\documentclass[a4paper,11pt]{article}
\pdfoutput=1 

\usepackage{jheppub} 

\usepackage[T1]{fontenc} 

\title{Holographic description of entanglement entropy of free CFT on torus}

\author[a]{Jun Tsujimura}

\affiliation[a]{Department of Physics,\\ 
Nagoya University, Chikusa, Nagoya 464-8602, Japan}

\emailAdd{tsujimura.jun.m5@s.mail.nagoya-u.ac.jp}

\abstract{
    The Ryu-Takayanagi conjecture provides a holographic description for the entanglement entropy for the strongly coupled holographic CFTs in the semi-classical limit. It proposes that the entanglement entropy is given by the area of the minimal homologous surface in the dual bulk. We show that the common terms of the entanglement entropy for the free massless fermions or bosons on a torus in the high-temperature expansion can be described by the sum of the signed area of extremal surfaces in the BTZ spacetime. The resulting EE and the corresponding bunch of the exgremal surfaces have preferable properties rather than those from the Ryu-Takayanagi conjecture.
}
\arxivnumber{}
\begin{document}
\maketitle
\flushbottom
\notoc

\section{Introduction}
\label{sec:1}

    The holographic principle of gravity proposes that the degrees of freedom of $d+1$-dimensional gravity is that of $d$-dimensional theory without gravity. This perspective is originated from the Beckenstein-Hawking formula that the black hole entropy is the area of its event horizon. The AdS/CFT correspondence \cite{Maldacena:1997re,Gubser:1998bc,Witten:1998qj} provides a toy model to investigate the holographic property of gravity. The Ryu-Takayanagi conjecture \cite{Ryu:2006bv,Ryu:2006ef} is a generalization of the Beckenstein-Hawking formula. It is believed to give a gravity dual of the entanglement entropy(EE) of the strongly coupled holographic CFTs in the semi-classical limit. Let $\rho_A$ denote the reduced density matrix of a space-like region $A$ in CFT, then the EE $S_A$ of a region $A$ is defined as 
\begin{align}\label{eq:def of EE}
    S_A = - \mathrm{Tr}_A \left(\rho_A \log \rho_A \right).
\end{align}
    For the semi-classical limit CFT, the Ryu-Takayanagi conjecture provides the EE of a region $A$ as
\begin{align}\label{eq:RT conjecture}
    S_A = \frac{\mathrm{Area\ of\ }\gamma_A}{4G},
\end{align}
    where $G$ is the gravitational constant and $\gamma_A$ is the global minimal surface in the dual bulk spacetime, that is homologous to the sub-system on the AdS boundary \cite{Headrick:2007km}. This conjecture has played a fundamental role in studying this issue and shed light on the relationship between the AdS/CFT correspondence and the information theory. It also brings many information-theoretical analyses \cite{Swingle:2009bg,Pastawski:2015qua,Lashkari:2013koa,Dong:2016eik} of the AdS/CFT correspondence. Thus, it is perhaps the most fundamental key to grasping the AdS/CFT correspondence.

    We will focus on the EE of one interval for free massless fermion or boson on the torus \cite{Azeyanagi:2007bj,Datta:2013hba}. We will see that the common terms in them can be described as the sum of the signed area of the extremal surfaces in the BTZ spacetime. Although a free CFT does not have the gravity dual, it is interesting to describe the EE of free CFTs in an holographic way, because it presents a geometrical point of view for the EE of free CFT and it allows us to compare the EEs between holographic CFTs and free CFTs. In addition, we will see that the configuration of the extremal surfaces and their signs has a geometrical consistency between the CFT side and the gravity side. Surprisingly, the resulting EE is smaller than the holographic EE given by the Ryu-Takayanagi formula and the corresponding bunch of the extremal surfaces has preferable holographic properties.

    The construction of this paper is as follows. In section~\ref{sec:2}, we will review the replica trick and discuss the geometrical consistency between the replica manifold of the CFT side and the gravity side. In section~\ref{sec:3}, we will see the extremal surfaces in the BTZ spacetime. In section~\ref{sec:4}, we will point out that the EE of one interval for free massless fermion on the torus is described by the sum of signed area of all the extremal surfaces that extend from the edge of the interval on the AdS boundary. Finally, section~\ref{sec:5} is the conclusion.

\section{Replica manifolds and deficit angle consistency}
\label{sec:2}

    We will investigate the geometry of the replica manifold and find that it has the conical singularities at the edge of a region. Eq.\eqref{eq:def of EE} can be rewritten by the replica trick \cite{Calabrese:2004eu, Calabrese:2009qy} as
\begin{align}\label{eq:replica trick}
    S_A = \lim_{n \to 1} S^{(n)}_A,\ \ \ S^{(n)}_A := \frac{1}{1-n} \log \mathrm{tr}_A \left(\rho_A^{\ n} \right).
\end{align}
    For convenience, we will consider the Renyi entropy $S^{(n)}_A$ in the following discussion. Let $Z$ be the partition function of a CFT on a manifold $\mathcal{B}$, and $Z_A^{(n)}$ be partition function of the same CFT on the replica manifold $\mathcal{B}_A^{(n)}$. Since $Z_A^{(n)}$ satisfies $\mathrm{tr}_A \left(\rho_A^{\ n} \right) =  Z_A^{(n)}/Z^n$, the Renyi entropy $S^{(n)}_A$ is written by the partition functions as 
\begin{align}\label{eq:replica trick2}
    S^{(n)}_A = \frac{1}{1-n} \left(  \log Z_A^{(n)} - n \log Z \right).
\end{align}
    For example, consider $\mathcal{B}=\mathbb{C}\cup \{ \infty \}$ and $A=[u,v]$. Instead of performing the path integral on the corresponding replica manifold $\mathcal{B}_A^{(n)}$, we can evaluate the Renyi entropy of $A$ as the following 2-point correlation function.
\begin{align}\label{eq:n replicated partition function}
    S^{(n)}_A = \frac{1}{1-n} \log \langle \mathcal{T}_n(u)\mathcal{\tilde{T}}_n(v) \rangle_{\mathcal{B}}
    = \frac{c(1+n)}{6 n} \log \frac{|u-v|}{\epsilon} + \mathrm{cnst}.
\end{align}
    where $\mathcal{T}_n$ and $\mathcal{\tilde{T}}_n$ are the primary twist operators with the conformal weight $h = c(n^2-1)/24n$.

    To investigate the geometry of the replica manifold $\mathcal{B}_A^{(n)}$ with $\mathcal{B}=\mathbb{C}\cup \{ \infty \}$ and $A=[u,v]$, consider the scale transformation of the Renyi entropy. For a moment, we will abbreviated the lower indices $A$ and specify the quantities defined on $\mathcal{B}^{(n)}$ by the upper index $(n)$. Applying the Ward-Takahashi identity for the scale transformation to eq.\eqref{eq:replica trick2}, we obtain 
\begin{align}\label{eq:scale trans for EE}
    \ell \frac{\partial}{\partial \ell} S^{(n)}
    &= \frac{2}{1-n} \left( \int_{\mathcal{B}^{(n)}} d^2x\, g^{(n)}_{\mu \nu} \frac{\delta}{\delta g^{(n)}_{\mu \nu}}  \log Z^{(n)} - n\, \int_{\mathcal{B}} d^2x\, g_{\mu \nu}\frac{\delta}{\delta g_{\mu \nu}} \log Z \right) \nonumber \\ 
    &= \frac{c}{24\pi(1-n)} \left(\int_{\mathcal{B}^{(n)}} d^2x\,\sqrt{\left|g^{(n)}\right|} R^{(n)} - n\, \int_{\mathcal{B}} d^2x\, \sqrt{\left|g\right|} R \right).
\end{align}
    where $\ell$ is a scale of the system, and $g_{\mu\nu},g$ and $R$ are the metric, the determinant of the metric and the Ricci scalar, respectively. Compered with eq.\eqref{eq:n replicated partition function}, the replica manifold $\mathcal{B}_A^{(n)}$ should be singular at $\partial A = \{u,v\}$ and we can consider the Ricci scalar as $R^{(n)} = -2\pi(n-1/n)\delta(t) ( \delta(x-u) + \delta(x-v) )$. Thus, there exists the conical deficit with angle $\Delta\phi = \pi(n-1/n)$ on the replica manifold at $\partial A$.

    We can also confirm that there exists conical singularity with angle $\Delta\phi = \pi(n-1/n)$ at $\partial A$ considering a CFT on $\mathbb{C}/\mathbb{Z}_N$ \cite{Nishioka:2006gr}, where $\mathbb{Z}_N = \mathbb{Z}/N$ and $N$ denotes an positive integer. Consider a free massless scalar field on $\mathbb{C}/\mathbb{Z}_N$ with the central charge $c$. Let $N=1/n$ and the sub-system $A = [0,\infty)$, then the partition function $Z_A^{(1/N)} $ is
\begin{align}\label{eq:partition function on Orbifold}
    \log Z_A^{(1/N)} 
    &= \frac{1}{N}\log Z + \frac{c}{N} \sum_{j=1}^{N-1} \int_{\mathbb{C}} dz d\bar{z}\, \delta(z-e^{\frac{2\pi i}{N} j} z) \delta(\bar{z}-e^{-\frac{2\pi i}{N} j} \bar{z}) \, \log \frac{\Lambda}{\epsilon} \nonumber \\
    &= \frac{1}{N}\log Z + \frac{c(N^2-1)}{12N} \int_{\mathbb{C}/\mathbb{Z}_N} d^2x\, \delta(t) \delta(x)\, \log \frac{\Lambda}{\epsilon}
\end{align}
    where $\Lambda$ and $\epsilon$ are the IR and UV cutoff lengths, respectively. Compared with eq.\eqref{eq:scale trans for EE}, we can understand that there exists conical singularity with angle $\Delta\phi = \pi(n-1/n)$ at a single edge of the interval $A$.

    Focus on the gravity dual of the above replica manifold $\mathcal{B}_A^{(n)}$. Consider the $2+1$-dimensional replica bulk manifold $\mathcal{M}_A^{(n)}$ of which boundary is the $1+1$ dimensional replica manifold $\mathcal{B}^{(n)}_A$ in the sense that $\partial \mathcal{M}_A^{(n)} = \mathcal{B}_A^{(n)}$. Since $\mathcal{B}_A^{(n)}$ has the conical singularities as discussed above, the replica bulk manifold $\mathcal{M}_A^{(n)}$ should have the canical singularities in the geometricary consistent manner. As the cosmic string with string tension $(n^2-1)/8 n G$ that makes the singularities with deficit angle $\pi(n-1/n)$ around it for $n \sim 1$, it is natural to consider that the replica bulk manifold $\mathcal{M}_A^{(n)}$ contains the cosmic string with string tension $(n^2-1)/8 n G$ \cite{Dong:2016fnf}. This seems the unique way of constructing the replica bulk manifold $\mathcal{M}_A^{(n)}$ consistently about the deficit angles between $\partial \mathcal{M}_A^{(n)}$ and $\mathcal{B}_A^{(n)}$. However, notice that we can consider for introducing the cosmic string with opposite signed string tension $-(n^2-1)/8n G$ and there are more configurations of cosmic strings satisfying this boundary condition than before. Actually, we will see that the Renyi entropy of free CFT with $n \sim 1$ is almost described as many cosmic strings with the both signed string tension satisfying the deficit angle consistency.

\section{Extremal surfaces in BTZ spacetime}
\label{sec:3}

\begin{figure}[ht]
    \centering
    \includegraphics[width=0.85\linewidth]{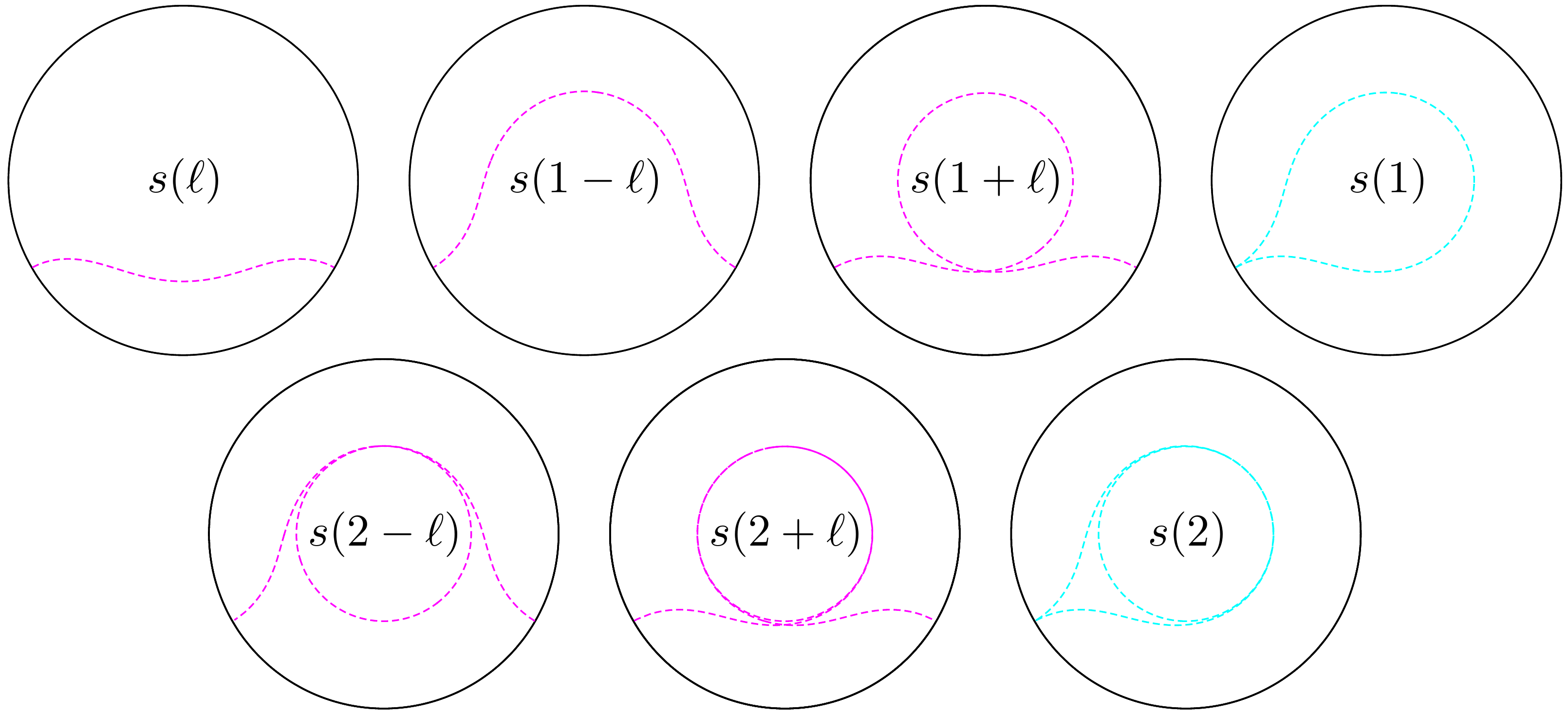}
    \caption{The extremal surfaces in the BTZ black hole spacetime with $M=1, L=1$. The red and blue lines represent the extremal surfaces described as eq.\eqref{eq:DefofAreaFunction} with $0 \le m \le 2$, that extend from $\theta_A = \pm \pi/3$ and $\theta_A = - \pi/3$, respectively. Each disk represents a time slice of the spacetime compactified for the radial direction. The outer circles represent the AdS boundary, and the black hole region is not written out. }
    \label{fig:ExtremalsInBTZ}
\end{figure}

    The cosmic strings for $n \sim 1$ as mentioned in the previous section behave as $1$-dimensional extremal surfaces. We will see the extremal surfaces in the BTZ spacetime and its areas. The metric of the BTZ spacetime can be described as follows.
\begin{align}\label{eq:BTZmetric}
    ds^2 = -\left(\frac{r^2}{L^2}-M\right) dt^2 + \left(\frac{r^2}{L^2}-M\right)^{-1} dr^2 + r^2 d\theta^2,
\end{align}
    where $M$ is the mass parameter of the black hole and $L$ is the AdS radius, and $t \in [-\infty,\infty]$, $r \in [\sqrt{M}L,\infty]$, $\theta \in [-\pi,\pi]$ and $\theta$-coordinate is periodic. As it is $2+1$-dimensional spacetime, we will derive space-like geodesics that extend from $\partial A:(t,r,\theta) = (0,\infty,\pm \theta_A)$ into the bulk on the $t=0$ time-slice. Minimizes the following length of a line. 
\begin{align}\label{line element in Example}
    \mathrm{Length} = \int_{-\theta_A}^{\theta_A} d\theta \sqrt{ \left(\frac{r^2}{L^2}-M\right)^{-1} \left( \frac{dr}{d\theta}\right)^2 + r^2 }.
\end{align}
    We will introduce the IR cut off scale in this integral later not to diverge for it. From this, a space-like geodesic on the $t=0$ time-slice in BTZ spacetime is described as 
\begin{align} \label{eq:extremal surface in BTZ}
    r(\theta) = \frac{\sqrt{M} L\, r_0\,\mathrm{sech}\left(\sqrt{M}\, \theta \right)}{\sqrt{M L^2 - r_0^2 \tanh^2 \left( \sqrt{M}\, \theta \right)}},
\end{align}
    where $r_0 := r(\theta=0)$. If this geodesic extends from each $(0,\infty,\pm \theta_A)$, the corresponding values are $r_0 = \sqrt{M} L \, \mathrm{coth} [\sqrt{M}(m \pi \pm \theta_A)]$, where $m$ is an integer satisfying $\pi m \pm \theta_A >0$. Each length of them are expressed as
\begin{align}\label{eq:finite-temp}
    \frac{\mathrm{Length}}{4G}
    = \frac{L}{2G} \log \left[ \frac{2\, r_\text{max}}{\sqrt{M}L } \sinh 
    \left( \sqrt{M}\,(m \pi \pm \theta_A) \right) \right]
\end{align}
    where we integrated from $r = r_0$ to $r = r_\text{max}$ to avoid the IR divergence, and we assumed $r_\text{max} \gg r_0$ and ignored the sub-leading terms. For the later discussion, consider the geodesics extend from only $(0,\infty, +\theta_A)$. We can immediately obtain them by replacing $\theta \to \theta +\theta_A$ in eq.\eqref{eq:extremal surface in BTZ}. In this case, $r_0 = \sqrt{M} L\, \mathrm{coth} [\sqrt{M} m \pi],\ m=1,2,\cdots$. Each length of them are also expressed as \eqref{eq:finite-temp} substituting $\theta_A=0$. Some of them are depicted in figure~\ref{fig:ExtremalsInBTZ}. Notice that $m$ represents the number of times that the corresponding geodesic passes through the opposite side of the black hole against the interval $[-\theta_A,\theta_A]$ on the AdS boundary. The minimal radius of the geodesic $r_0$ approaches to the black hole horizon radius $\sqrt{M} L$ taking $m \to \infty$. 

    The BTZ black hole spacetime is considered as the dual gravity spacetime of $1+1$ dimensional CFT on torus in the AdS/CFT correspondence. Let us consider CFT defined on $\mathrm{S}^1$ of which circumference is $C$ with the UV cutoff length $\epsilon$ and the inverse temperature $\beta$. The bulk parameters are translated into the above CFT quantities as follows. The black hole mass and the IR cut off are $\beta/C=1/\sqrt{M}$ and $\epsilon= L /(2\pi r_\text{max})$. The Brown-Henneaux formula \cite{Brown:1986nw} translates $G$ into the central charge $c$ of CFT as $c =3L/(2G)$. The length of an interval is $\ell := C\theta_A/\pi \in [0,C]$. Then, we can describe eq.\eqref{eq:finite-temp} as follows.
\begin{align} \label{eq:DefofAreaFunction}
    s(m C \pm \ell) := \frac{c}{3} \log \left[ \frac{\beta}{\pi C \epsilon} \sinh 
    \left\{ \frac{\pi(mC \pm \ell)}{\beta} \right\} \right].
\end{align}
    In the next section, we will see that the EE of one interval for the free massless field on a torus is almost described by an appropriate sum of $\pm s(m C \pm \ell)$.

\section{Holographic description for EE on torus}
\label{sec:4}

\begin{figure}[ht]
    \centering
    \includegraphics[width=0.74\linewidth]{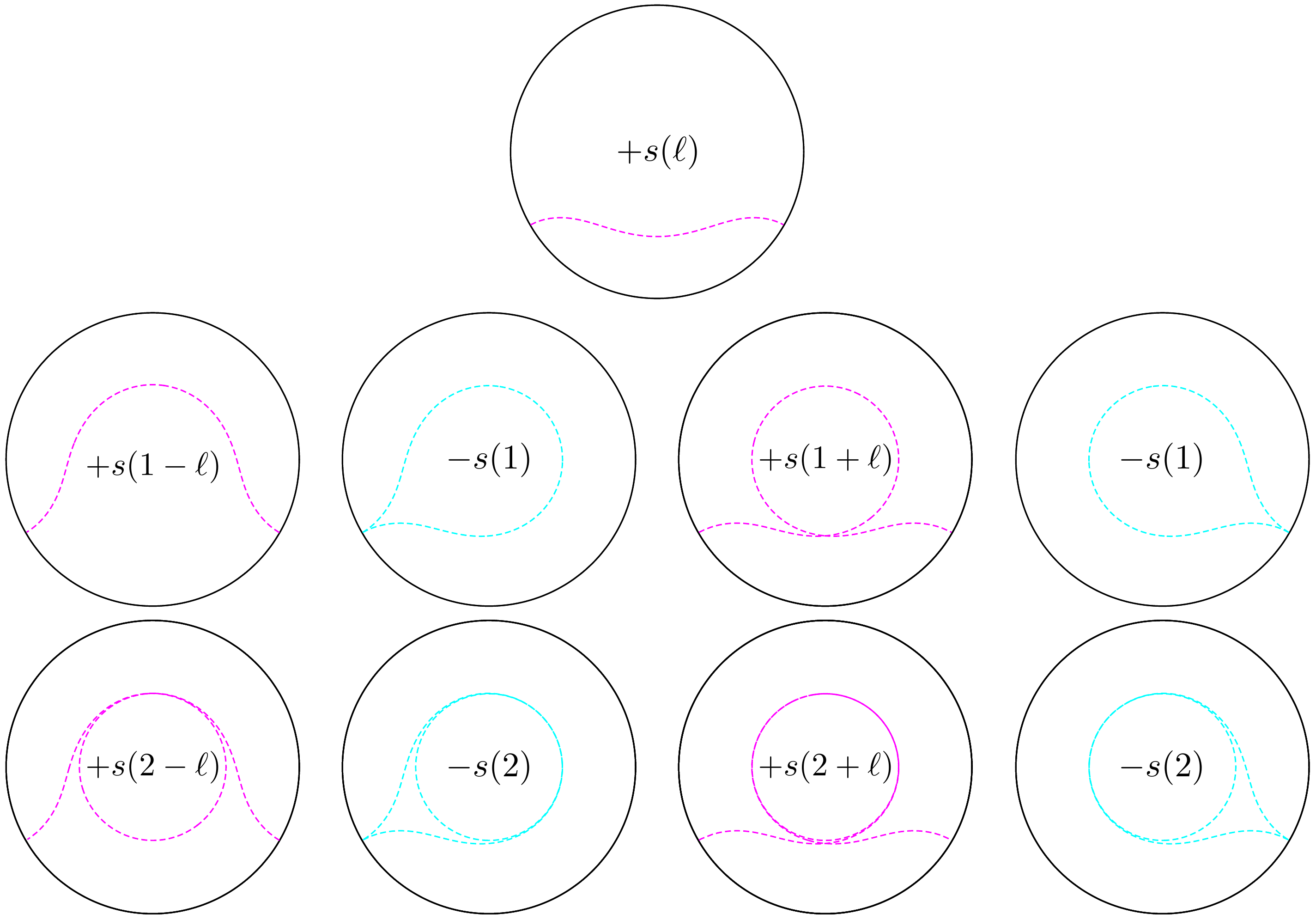}
    \caption{The diagrams of the extremal surfaces which contribute to the EE of a single interval in the BTZ black hole spacetime. The function $s$ defined as eq.\eqref{eq:DefofAreaFunction} denotes the area$/4G$ of each surface. The total winding number of the extremal surfaces with the same $m$ is $0$.}
    \label{fig:ThesurfacesInBTZ}
\end{figure}

    Consider $1+1$-dimensional CFT on the circle of which circumference $C=1$ at the inverse temperature $\beta$. The common term of the EE of an interval $A = [-\ell/2, \ell/2],\, \ell \in (\epsilon,1-\epsilon)] $ on this system in the high-temperature expansion is as follows \cite{Azeyanagi:2007bj,Datta:2013hba}.
\begin{align}\label{eq:EE of finite circle}
    S_A
    &= \frac{c}{3} \left\{ \log\left[ \frac{\beta}{\pi\epsilon} \sinh\left(\frac{\pi\ell}{\beta}\right)\right]
    + \sum_{m=1}^{\infty}\log \frac{(1-e^{2\pi \ell/\beta}e^{-2\pi m/\beta})(1-e^{-2\pi \ell/\beta}e^{-2\pi m/\beta})}{(1-e^{-2\pi m/\beta})^2} \right\} \nonumber\\
    &=\frac{c}{3} \log\left[ \frac{\beta}{\pi\epsilon} \sinh\left(\frac{\pi\ell}{\beta}\right)\right]
    + \frac{c}{3} \sum_{m=1}^{\infty}\log \frac{\sinh \left[\frac{\pi}{\beta}(m-\ell)\right]\sinh \left[\frac{\pi}{\beta}(m+\ell)\right]}{\sinh^2 \left(\frac{\pi m}{\beta}\right)}
\end{align}
    This EE is described as the following sum of signed length of extremal surfaces from eq.\eqref{eq:DefofAreaFunction}. 
\begin{align}\label{eq:BTZ}
    S_A = s(\ell) + \sum_{m=1}^{\infty} \left[ s(m - \ell) + s(m + \ell) - 2 s(m) \right]
\end{align}
    The corresponding extremal surfaces are depicted as fig.~\ref{fig:ThesurfacesInBTZ}.
    From an algebraic viewpoint, we cannot determine the configurations of each surface. In particular, the surfaces corresponding to $-s(1)$ seem to need not extend from each $\pm \ell/2$. However, from the conical deficit angle consistency as we discussed in section \ref{sec:2}, it is natural to describe the surface configuration as depicted in fig.~\ref{fig:ThesurfacesInBTZ}.

    In what follows, we will comment on this holographic description for free field EE compared with the Ryu-Takayanagi conjecture in the BTZ black hole spacetime. The Ryu-Takayanagi conjecture predicts the corresponding EE as
\begin{align}\label{eq:EEfromRT}
    S_A = \mathrm{min}\left\{ s(\ell),\ s(1-\ell) + S_{BH} \right\},
\end{align}
    where $S_{BH} = c\pi/3\beta$ corresponds with the black hole entropy. We will focus on a few differences between them on geometrical aspects of the configuration of the extremal surfaces corresponding to them. First, the most striking difference between eq.\eqref{eq:EEfromRT} and eq.\eqref{eq:BTZ} is the region where the surfaces can sweep. The homologous minimal surface corresponding with the Ryu-Takayanagi conjecture has the entanglement shadow region and the plateaux problem \cite{Freivogel:2014lja, Hubeny:2013gta}. Since any homologous minimal surface cannot reach the bulk region near by the black hole horizon, the holographic EE is independent of the property of such a region. On the other hand, since the surfaces corresponding with eq.\eqref{eq:BTZ} can sweep all the bulk region outside the black hole horizon, the plateaux problem does not happen. In the same way, since in the higher dimensional Schwarzschild-AdS black hole there are surfaces that can approach the black hole horizon infinitesimally, the plateaux problem may not happen for free CFT as well.

    Second, we will see the relation between surfaces of eq.\eqref{eq:BTZ} and the black hole horizon. Although eq.\eqref{eq:BTZ} seems not to include the black hole horizon explicitly, if $A = 1-\delta,\, \epsilon < \delta \ll 1$, eq.\eqref{eq:BTZ} becomes 
\begin{align}
    S(1-\delta) &= \frac{c}{3} \log\left[ \frac{\beta}{\pi\epsilon} \sinh\left(\frac{\pi \delta}{\beta}\right)\right]
    + \frac{c}{3} \lim_{m \to \infty} \log \frac{\sinh \left(\frac{\pi}{\beta}(m+1)\right)}{\sinh \left(\frac{\pi m}{\beta}\right)} + O(\delta) \nonumber \\
    &= \frac{c}{3} \log\left[ \frac{\beta}{\pi\epsilon} \sinh\left(\frac{\pi \delta}{\beta}\right)\right] + \frac{c \pi}{3 \beta}.
    \label{eq:BTZ large interval}
\end{align}
    Taking $\delta, \epsilon \to 0$, the Araki-Lieb inequality is saturated and the difference of the EEs corresponds to the black hole entropy $S_{BH}$:
\begin{align}
    \lim_{\delta,\epsilon \to 0} S(1-\delta)- S(\delta) = \frac{c \pi}{3 \beta} = S_{BH}.
    \label{eq:Araki-Lieb}
\end{align}
    The black hole horizon emerges as a result of the difference of the surface wrapped $m+1$ times around the black hole and one wrapped $m$ times taking $m$ to the infinity. Both sides of eq.\eqref{eq:Araki-Lieb} are equivalent as the surfaces not just as the value.

    Finally, we should consider that eq.\eqref{eq:BTZ} may describe the holographic EE for the holographic CFTs. The EE given by eq.\eqref{eq:BTZ} is smaller than that from the Ryu-Takayanagi formula eq.\eqref{eq:EEfromRT}. To consider the hamologous condition, pay attention for the topology of the surface configuration in fig.~\ref{fig:ThesurfacesInBTZ}. The total winding number of the surfaces for $m \ge 1$ around the black hole is $0$. Thus, when we regard all the surfaces in fig.~\ref{fig:ThesurfacesInBTZ} as a single surface, it is homotopy equivalence to the sub-region $A$ on the AdS boundary. Therefore, eq.\eqref{eq:BTZ} may be the true holographic EE.

\section{Conclusion}
\label{sec:5}

    We provided a holographic description of the entanglement entropy given by eq.\eqref{eq:EE of finite circle} as the sum of the signed areas of the extremal surfaces satisfying the deficit angle consistency. It has preferable properties for the HEE in the following reasons. First, it does not have the entanglement shadow region that the Ryu-Takayanagi conjecture has. Second, the resulting holographic EE eq.\eqref{eq:BTZ} gives smaller EE than that from the Ryu-Takayanagi conjecture, and the bunch of the surfaces corresponding with eq.\eqref{eq:BTZ} is homotopy equivalent to the sub-system on the AdS boundary. Thus, eq.\eqref{eq:BTZ} can be a candidate of the holographic EE for the holographic CFT.

\bibliographystyle{JHEP}
\bibliography{HolographicDescriptionOfEntanglementEntropyOfFreeCFTonTorus}
 
\end{document}